\documentclass{ws-p8-50x6-00}
\usepackage{amsmath}

\newcommand{\ben}{\begin{enumerate}}
\newcommand{\een}{\end{enumerate}}
\newcommand{\bit}{\begin{itemize}}
\newcommand{\eit}{\end{itemize}}
\newcommand{\la}[1]{\label{#1}}

\newcommand{\eq}[1]{eq.\thinspace(\ref{#1})}

\newcommand{\half}{\frac{1}{2}}

\newcommand{\hf}{{\textstyle \frac{1}{2}}}

\newcommand{\uu}{\mathord{\uparrow}}
\newcommand{\dd}{\mathord{\downarrow}}
\newcommand{\fract}[2]{{\textstyle\frac{#1}{#2}}}
\newcommand\undertilde[1]{\smash{\mathop{#1}\limits_{\widetilde{}}}}
\input BoxedEPS
\SetRokickiEPSFSpecial  
\HideDisplacementBoxes

\makeatletter
   \def\@oddfoot{\lower2pc\hbox{\centerline{\reset@font \small  R.L. Jaffe \it --- 
            Color, Spin, and Flavor-Dependent Forces in QCD}\hfill \llap{\bf \thepage}}}
\makeatother

\begin{document}

\title{Color, Spin, and Flavor-Dependent Forces \\
in Quantum Chromodynamics}

\author{R.L.~Jaffe}

\address{Center for Theoretical Physics and Department of Physics \\
Laboratory for Nuclear Physics
Massachusetts Institute of Technology\\
Cambridge, Massachusetts 02139\\E-mail: jaffe@mit.edu\\[1ex]
{\small In Memoriam  Kenneth A. Johnson
1931--1999}}

\maketitle

\abstracts{
A simple generalization of the Breit Interaction explains many 
qualitative features of the spectrum of hadrons.\hfill
hep-ph/0001123\qquad  {MIT CTP \# 2938}
}

\thispagestyle{empty}

\section{Introduction}
\label{section1}
I did not know Gregory Breit, but two of his closest associates 
influenced me deeply.  I did research in Gerry Brown's group as
an undergraduate at Princeton.  Gerry welcomed young students into his
research family and gave us wonderful problems to work on.  He taught
us that complicated problems often have simple answers -- something
that has proved true in particle physics over the past quarter
century.  Vernon Hughes invented deep inelastic spin physics in the
early 1970's and has championed it relentlessly ever since. 
He and his collaborators have literally rewritten the book on the
quark and gluon structure of the nucleon.  It is a pleasure to speak
at a symposium celebrating their teacher, Gregory Breit.

My talk will be largely pedagogical.  I would like to show how a
generalization of the ``Breit Interaction'' can account for some of
the regularities of hadron physics.  Although the basic ideas
described here date back to the 1970's, they have not been presented
quite this way before, and some developments are still at the
forefront of modern research in QCD\null. This subject was a special
favorite of my friend and collaborator, Ken Johnson, who died this
past winter.  Ken had many friends at Yale.  He would have liked to
hear this story, so I dedicate my talk to his memory.

It is hard to make definite statements about hadrons made of the light
$u$, $d$, and $s$ quarks.  The nonperturbative regime is too
complicated.  It may never be solved to our satisfaction except on a
computer.  Nevertheless, the spectrum and interactions of baryons and
mesons display remarkable regularities, which correlate with simple
symmetry properties of the fundamental quark/gluon interactions.  The
role of models in QCD is to  build simple physical pictures that connect the
phenomenological regularities with the underlying structure.

The QCD ``Breit Interaction'' is the spin-dependent part of one-gluon
exchange between light quarks in the lowest state of some unspecified
mean field.  It is summarized by an effective Hamiltonian acting on
the quarks' spin and color indices,
\begin{equation}
        {\cal H}_{\rm eff} \propto\, -\!\sum_{i\ne j}
        {\undertilde{\lambda}}\,_{i}\cdot {\undertilde{\lambda}}\,_{j}
        \vec{\sigma}_{i}\cdot\vec{\sigma}_{j} 
        \la{0.1}
\end{equation}
where $\vec\sigma_{i}$ and ${\undertilde\lambda}\,_{i}$ are the spin
and color operators of the $i^{\rm th}$ quark.  The spin operators are
represented by the three $2\times 2$ Pauli matrices, normalized to
$\mathop{\rm Tr}(\sigma_{i}^{k})^{2}=2$ for $k=1,2,3$, and the color
operators are represented by the eight $3 \times 3$ Gell-Mann
matrices, normalized the same way, $\mathop{\rm Tr}(\lambda_{i}^{a})^{2}=2$
for $a=1,\ldots8$.  The sum over $i$ and $j$ extends over all quark
pairs.  For the moment, I ignore antiquarks.  I also ignore quark mass
differences.  In reality the $u$ and $d$  masses are small enough
relative to the natural scale of QCD that we can neglect them.  The
$s$ quark is heavier.  It is a reasonable first approximation to
ignore its mass as well.  The $c$, $b$, and $t$ quarks are too heavy,
and cannot be treated this way.  The space-time dependence of ${\cal
H}_{\rm eff}$ is not well understood, but in this approximation it is
universal, and need not concern us much.

\begin{figure}[t]
\begin{center}
\BoxedEPSF{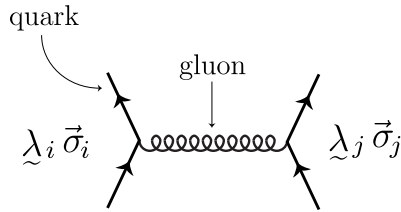 scaled 1200} 
\end{center}
 \caption{One-gluon exchange between quarks.\vspace*{-1pc}}
\label{fig1}
 \end{figure}

${\cal H}_{\rm eff}$ can be read off the Feynman diagram for one-gluon
exchange between quarks; see Fig.~(\ref{fig1}).  At short distances
where QCD is weakly coupled, we can trust perturbation theory, and
one-gluon exchange should dominate.  However, at typical hadronic
distance scales QCD is strongly coupled.  Still, there is reason to
take the qualitative predictions which follow from ${\cal H}_{\rm
eff}$ seriously.  Once the long range, spin-average, confining
interactions in QCD have been integrated out, the resulting confining,
bag-like mean field acts as an infrared cutoff, reducing the strength
of the remaining QCD effects.  Most phenomenological models of QCD --
the Bag Model, the nonrelativistic quark model, and other quark
models in particular -- use this picture successfully.  The absence of
strong renormalization (higher twist effects) in deep inelastic
scattering offers phenomenological support for a picture of hadrons
where perturbation theory is qualitatively reliable once confinement has been
implemented.

A couple of further notes on the form of \eq{0.1}: First, the spin-averaged piece
of one-gluon exchange has been set aside.  It figures in the dynamics of
confinement but not in the spectroscopy considered here.  Second, the tensor
and spin-orbit interactions generated by one-gluon exchange average to zero in
the lowest quark state.  Third, the appearance of $\vec\sigma$ matrices in
\eq{0.1} does not mean the analysis is nonrelativistic.  For light quarks this
would be an unacceptable restriction.  The Dirac $\vec\alpha$ matrices which
appear in the relativistic quark currents reduce to $\vec \sigma$
matrices in the lowest orbital.  For obvious reasons the interaction
of \eq{0.1} is known as the ``colorspin'' or ``color-magnetic''
interaction of QCD\null.

${\cal H}_{\rm eff}$ was first introduced by De Rujula, Georgi, and
Glashow in their pioneering paper on hadron spectroscopy in
QCD\null.\cite{DeRujula:1975ge} Many of the spectroscopic results I will
discuss were developed by them or by our group at MIT in the
mid-1970's.\cite{DeGrand:1975cf,Jaffe:1977ig,Jaffe:1977yi} The subject
of the scalar mesons has been revitalized recently by Schechter and
his collaborators.\cite{Black:1998wt} The possible role of ${\cal
H}_{\rm eff}$ in quark matter has been the subject of much recent
activity starting with the fundamental work of Alford, Rajagopal, and
Wilczek.\cite{Alford:1997zt}

My talk is organized as follows: In Section~\ref{section2} I review the basic
symmetry structure of ${\cal H}_{\rm eff}$.   In Section~\ref{section3} I look at
some properties of the baryons: the octet-decuplet splitting, the
$\Lambda$--$\Sigma$ splitting, and the pattern of excitations.  
Section~\ref{section4} is devoted to mesons: first the pseudoscalar-vector
splittings, next a remark on the absence of exotics, and finally a reevaluation of
the $J^{PC}=0^{++}$ mesons.  In Section~\ref{section5} I return to symmetry and
extract a simple rule for the ground state of the $Q^{N}$ configuration -- the
rule of flavor antisymmetry.  In Section~\ref{section6}, I apply it to baryons
($Q^{3}$) and dibaryons ($Q^{6}$).  Finally, in Section~\ref{section7} I give a
very brief introduction to the effects of
${\cal H}_{\rm eff}$ in quark matter -- condensates, superconductivity, and 
unusual patterns of symmetry breaking.

\section{Basics:  Regularities of the $\mathbf{Q}$--$\mathbf{Q}$ interaction}
\label{section2}
It is not necessary to use much mathematics to understand the 
implications of \eq{0.1} for the simplest case of two quarks.  Most of 
what I need can be done merely by rewriting it in terms of color, 
spin, and flavor ``exchange operators''.  The spin exchange operator, 
$P^{S}_{12}$, is defined by
\begin{equation}
	P^{S}_{12} = \hf +\hf\vec\sigma_{1}\cdot\vec\sigma_{2}\ .
	\la{1.1}
\end{equation}
Two spin-$\half$ particles may be coupled to a triplet of spin-$1$
states, $|3_{\rm s}\rangle\equiv\{|\uu\uu\nobreak\rangle,\frac{1}{\sqrt{2}}
|\uu\dd+\dd\uu\rangle,|\dd\dd\rangle\}$, which are symmetric under
spin exchange, or to a singlet spin-$0$ state, $|1_{\rm s}\rangle \equiv
\frac{1}{\sqrt{2}}|\uu\dd-\dd\uu\rangle$, which is antisymmetric under
spin exchange.\footnote{To avoid confusion, we label all states by
their degeneracy.  Thus the spin-$1$ states are $|3_{\rm s}\rangle$ and
the spin-$0$ states are $|1_{\rm s}\rangle$.} The eigenvalues of
$\vec\sigma_{1}\cdot\vec\sigma_{2}$ are $+1$ and $-3$ in the spin-$1$
and spin-$0$ states respectively, so $P^{S}|3_{\rm s}\rangle =
+|3_{\rm s}\rangle$, $P^{S}|1_{\rm s}\rangle = -|1_{\rm s}\rangle$.  Thus $P^{S}$
has the desired property that it gives $\pm 1$ on states which are
symmetric/antisymmetric under spin exchange.

Color and flavor states are both classified by $SU(3)$, so color and
flavor exchange operators have the same form.  To be specific,
consider flavor.  The quark labels are $u$, $d$, and $s$.  I need a
notation for the set of Gell-Mann matrices associated with flavor --
denote them by $\{\undertilde{\beta\null}\}$ to distinguish them from the
$\{\undertilde\lambda\}$ used for color.

\begin{figure}[t]
\begin{center}
\BoxedEPSF{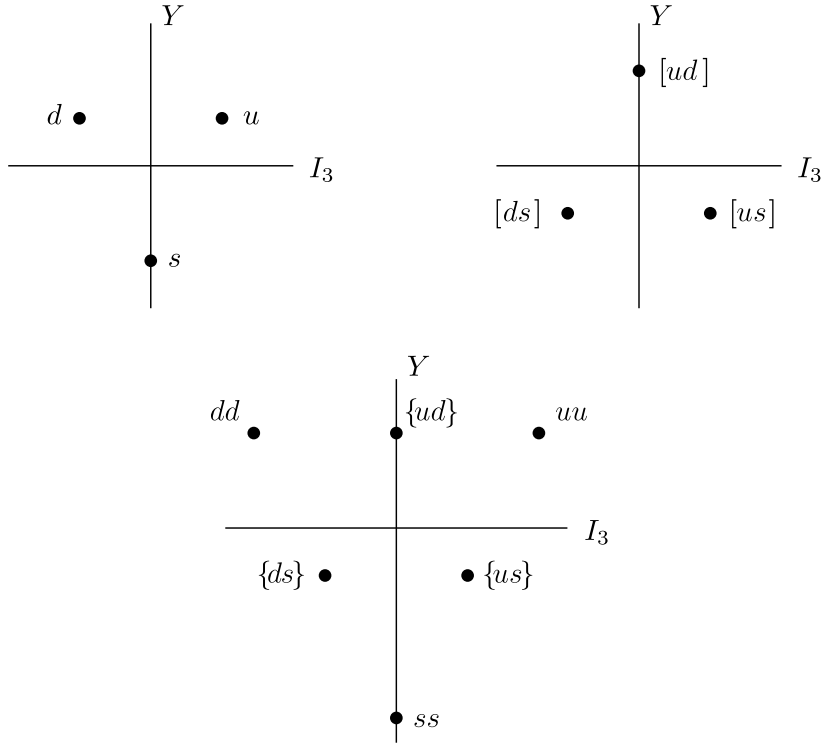 scaled 1100} 
\end{center}
 \caption{$SU(3)_{\rm f}$ weight diagram for the fundamental $3_{\rm f}$,
antisymmetric diquark, $\bar 3_{\rm f}$, and symmetric diquark, $6_{\rm f}$
representations.\vspace*{-1pc}}
\label{fig2}
 \end{figure}

There are nine flavor states of two quarks.  Six are symmetric:
$|6_{\rm f}\rangle \equiv \{ |uu\rangle,|dd\rangle,|ss\rangle,
\frac{1}{\sqrt{2}}|ud+du\rangle, \frac{1}{\sqrt{2}}|ds+sd\rangle,
\frac{1}{\sqrt{2}}|su+us\rangle\}$.  Three are antisymmetric: $|\bar
3_{\rm f}\rangle\equiv \{ \frac{1}{\sqrt{2}}|ud-du\rangle,
\frac{1}{\sqrt{2}}|ds-sd\rangle, \frac{1}{\sqrt{2}}|su-us\rangle\}$. 
These states can be visualized most easily in the usual $SU(3)_{\rm f}$
weight diagram in which the third component of isospin ($I_{3}=
\half(n_{u}-n_{d})$) is the $x$-axis and hypercharge
($Y=\fract{1}{3}(n_{u}+n_{d}-2n_{\rm s}$)) is the $y$-axis.  The original
quark triplet forms a triangle, point down, the $3_{\rm f}$.  The
antisymmetric set of two quark states forms a triangle, point up, hence
the notation $\bar 3_{\rm f}$.  The symmetric set of two quark states form
a larger triangle, point down, the $6_{\rm f}$.  All three representations
are shown in Fig.~(\ref{fig2}).

A simple exercise with the Gell-Mann matrices lead to the exchange 
operator for flavor,
\begin{equation}
	P^{F}_{12} = \fract{1}{3} + \hf{\undertilde\beta}\,_{1}\cdot
	{\undertilde{\beta}}\,_{2} 
	\la{1.3}
\end{equation}
with the desired property that $P^{F} |6_{\rm f}\rangle =
+|6_{\rm f}\rangle$ and $P^{F} |\bar 3_{\rm f}\rangle = -|\bar
3_{\rm f}\rangle$.
Since color has the same structure as flavor:
\begin{equation}
	P^{C}_{12} = \fract{1}{3} + \hf{\undertilde\lambda}\,_{1}\cdot
	{\undertilde\lambda}\,_{2} 
	\la{1.5}
\end{equation}
and $P^{C} |6_{\rm c}\rangle = +|6_{\rm c}\rangle$,
$P^{C} |\bar 3_{\rm c}\rangle = -|\bar 3_{\rm c}\rangle$.

Armed with eqs.\thinspace(\ref{1.1}), (\ref{1.3}), and (\ref{1.5}), we can 
rewrite the interaction between two quarks in terms of exchange 
operators,
\begin{equation}
   	{\cal H}_{\rm eff} \propto -  {\undertilde{\lambda}}\,_{1}\cdot
   	{\undertilde{\lambda}}\,_{2} \vec{\sigma}_{1}\cdot\vec{\sigma}_{2} =
   	-4P^{C}_{12}P^{S}_{12}+\fract{4}{3}P^{S}_{12} 
   	+2P^{C}_{12}-\fract{2}{3}\, .
	\la{1.7}
\end{equation}

The next step exposes the reason for all this algebra:  two quarks in 
the same orbital are
\emph{symmetric} under space exchange.  Since the quarks are 
fermions, the wavefunction must be \emph{antisymmetric} under the 
simultaneous exchange of the remaining labels:  spin, color, and 
flavor.  In terms of exchange operators,
\begin{eqnarray}
	P^{C}_{12}P^{S}_{12}P^{F}_{12} &=& -1 \qquad \hbox{or}
	\nonumber\\
	P^{C}_{12}P^{S}_{12} &=& - P^{F}_{12}\, .
	\la{1.8}
\end{eqnarray}
So \eq{1.7} can be rewritten as
\begin{equation}
	{\cal H}_{\rm eff}= 4P^{F}_{12}+\fract{4}{3}P^{S}_{12}
	+2P^{C}_{12}-\fract{2}{3}\, .
	\la{1.9}
\end{equation}

Lo and behold:  The dominant piece of the colorspin force (weighted 
by~$4$) is a \emph{flavor}-exchange interaction.  \emph{This is why 
colorspin has significant implications for flavor-dependent mass 
splittings}.

There are four totally antisymmetric configurations of 
flavor$\times$spin$\times$color.  They and their associated 
colorspin interaction strengths are listed in \hbox{Table~\ref{tab1}}.
\begin{table}[t]
\caption{Flavor, spin, and color states of two quarks,  and the eigenvalue of the
QCD  Breit Interaction of \eq{0.1}.\label{tab1}}
\begin{center}
\footnotesize
\begin{tabular}{|ccc|c|}
\multicolumn{1}{c}{Flavor} & \multicolumn{1}{c}{Spin}
 & \multicolumn{1}{c}{Color} & \multicolumn{1}{c}{$\Delta E$}\\
\hline
$\bar3(A)$ & $1(A)$ & $\bar 3(A)$ & $-8$\rule{0pt}{3ex}\\[1ex]
\hline
$\bar3(A)$ & $3(S)$ & $6(S)$ & $-4/3$\rule{0pt}{3ex}\\[1ex]
\hline
$6(S)$ & $3(S)$ & $\bar 3(A)$ & $8/3$\rule{0pt}{3ex}\\[1ex]
\hline
$6(S)$ & $1(A)$ & $6(S)$ & $4$\rule{0pt}{3ex}\\[1ex]
\hline
\end{tabular}
\end{center}
\end{table}
From \eq{1.9} it is clear that the configuration separately
antisymmetric in flavor, spin, and color, is the most attractive channel. 
The other three configurations are symmetric in at least two indices,
leading to dramatically less binding.  The dominance of a single $qq$
configuration, $|\bar 3_{\rm f}1_{\rm s}\bar 3_{\rm c}\rangle$, is the basic
observation behind colorspin phenomenology.  Colorspin favors
antisymmetric (and therefore low-dimension) representations of the
flavor and spin symmetry groups.  Several applications will bear this
out.

\section{Baryons}
\label{section3}
\subsection{The Octet-Decuplet Mass Difference}

\begin{figure}[t]
\begin{center}
\BoxedEPSF{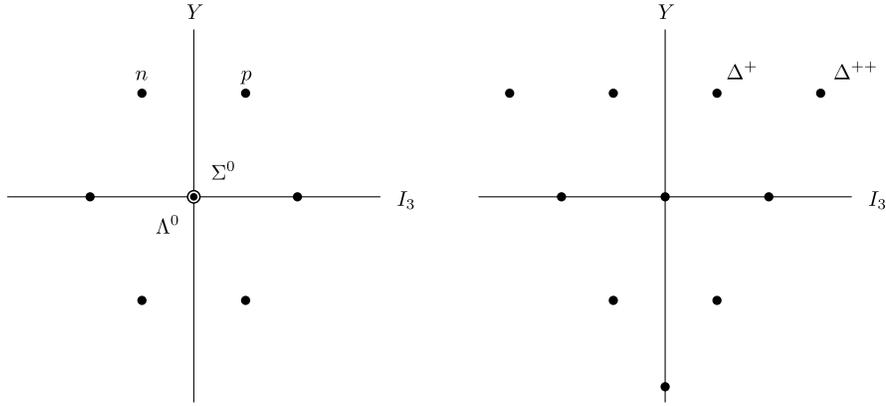 scaled 800} 
\end{center}
 \caption{$SU(3)_{\rm f}$ weight diagram for the baryon octet and decuplet. 
Some states  discussed in the text are labeled.\vspace*{-1pc}}
\label{fig3}
 \end{figure}

The lightest baryons composed predominantly of $u$, $d$, and $s$
quarks can be grouped into an $SU(3)_{\rm f}$ octet with spin-$\half$ and an
$SU(3)_{\rm f}$ decuplet with spin-$\fract{3}{2}$.  The $SU(3)_{\rm f}$ weight
diagrams in terms of $I_{3}$ and $Y$ are shown in Fig.~(\ref{fig3}).  States
with the same $I_{3}$ and $Y$ contain the same valence quarks.  Thus
the $\Delta^{+}$ and $p$ must both contain $uud$ in addition to any
$\bar qq$ pairs and gluons.  For all pairs like the $p-\Delta^{+}$, the
state in the decuplet is roughly $200$--$250$~MeV heavier than the
equivalent state in the octet.  The \emph{sign} of this effect is
easily explained from the symmetry properties of the colorspin force:
Since baryons are color singlets, each quark pair must be coupled to a
color $\bar 3_{\rm c}$.\footnote{So a color singlet can be made when the
third quark is included:  $\bar 3\otimes 3=1\oplus 8$, whereas
$6\otimes 3 = 8\oplus 10$.} Because the decuplet has
spin-$\fract{3}{2}$, each quark pair must be coupled to a spin triplet. 
Referring to Table 1, it appears that the $\bar 3_{\rm c}3_{\rm s}$ colorspin
configuration has positive interaction energy.  The spin coupling of
the quarks in the octet is more complicated, but one thing is clear:
at least the coupling of quark pairs to spin-$0$ is allowed. 
Therefore a state in the octet must be lighter than state with the
same quark content in the decuplet.

\subsection{The $\Lambda$--$\,\Sigma$ Splitting}

The lightest neutral strange baryons, the $\Lambda^{0}(1116)$ and the 
$\Sigma^{0} (1192)$ have the same quark content, $uds$.  They differ 
in mass by $76$~MeV\null.  Before QCD this mass difference puzzled 
theorists who tried to understand the quark structure of hadrons.  The 
puzzle was resolved by DeRujula, Georgi, and Glashow and led to much 
work on QCD inspired quark models.\cite{DeRujula:1975ge}  The 
$\Lambda^{0}$ and the $\Sigma^{0}$ have the same quarks and the same spin.  
They differ in isospin:
\begin{equation}
	| \Sigma^{0}\rangle \sim |(ud)^{I=1}s\rangle \qquad \
	|\Lambda^{0}\rangle \sim |(ud)^{I=0}s\rangle \ .
	\la{2.1}
\end{equation}
Colorspin, like all spin-dependent effects in gauge theories, is a
relativistic effect.  For heavy quarks the coefficient we have
suppressed in front of ${\cal H}_{\rm eff}$ goes like $1/m_{q}^{2}$. 
It is reasonable to assume that colorspin forces weaken with
increasing quark mass.  The $u$ and $d$ quarks are much lighter than
the $s$ quark, so the colorspin force between a $ud$ pair is stronger
than that between a $us$ or $ds$ pair.  The $ud$ pair in the
$\Lambda^{0}$ is in the $I=0$ or $\bar 3_{\rm f}$ state.  In the
$\Sigma^{0}$, the $ud$ pair is in the $I=1$ or $6_{\rm f}$ state.  The
$ud$ colorspin contribution to the mass is therefore attractive in the
$\Lambda^{0}$, repulsive in the $\Sigma^{0}$.  Since this dominates
over $ds$ and $us$, the $\Lambda^{0}$ must be lighter than the
$\Sigma^{0}$.  The magnitude of the $\Sigma^{0}-\Lambda^{0}$ mass
difference depends on the strength of the colorspin interaction (fixed
by the decuplet-octet splitting) and the quark mass differences (fixed
by $SU(3)_{\rm f}$ violating mass differences).  The numerical value is
roughly consistent with the $76$~MeV observed.

\subsection{Excited Baryon Spectroscopy}

Tables of the masses and properties of excited baryons fill
books.\cite{Caso:1998} Long ago, Dalitz and collaborators\cite{Dalitz}
showed that the fundamental structure of this spectrum could be
understood in a ``Symmetric Quark Model'': Populate
spin$\times$flavor$\times$space symmetric states of three quarks
(color antisymmetry takes care of Fermi statistics) in the modes of
some simple central potential.  The lightest (positive parity) baryons
have all quarks in the lowest orbital.  To obtain negative parity
excitations, promote one quark to the first ($L=1$) excited orbital. 
To make positive parity excitations, promote two quarks to the first
level or one quark to the second. 

\begin{figure}[t]
\begin{center}
\BoxedEPSF{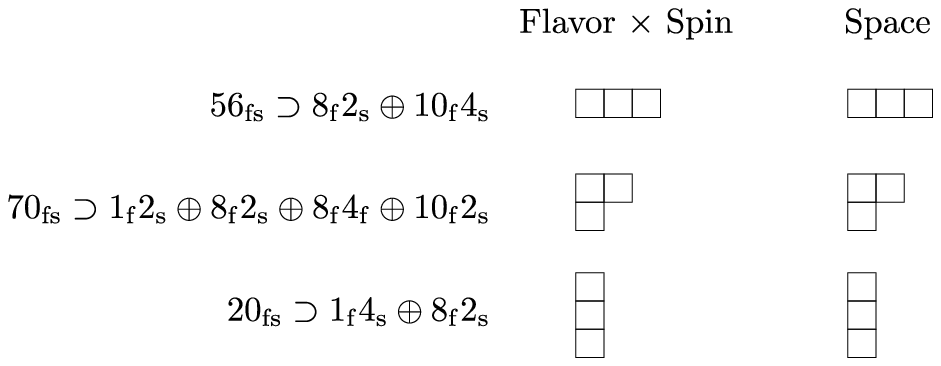} 
\end{center}
 \caption{States of three quarks. $SU(6)_{\rm fs}$ multiplets are shown along
with  their $SU(3)_{\rm f}\otimes SU(2)_{\rm s}$ content.  The multiplets are
represented by  Young Diagrams.  Boxes stacked vertically represent
antisymmetrized  quarks, those stacked horizontally are symmetrized.  To make
states  symmetric under flavor$\times$spin$\times$space, quarks must have
the same  permutation symmetry under flavor$\times$spin and space.  Hence
the  identical Young Diagrams.\vspace*{-1pc}}
\label{fig4}
 \end{figure}

Dalitz's model contains the G\"ursey's famous $SU(6)_{\rm fs}$
symmetry\footnote{I use the subscript ${\rm fs}$ in order
not to mix it up with another $SU(6)_{\rm cs}$ built of color$\times$spin, which
will be useful later.} built out of 
$SU(3)_{\rm f}\otimes SU(2)_{\rm s}$.\cite{Gursey} Baryon multiplets form
representations of $SU(6)_{\rm fs}$, which in turn contain $SU(3)_{\rm f}$
multiplets of definite spin.  The three $SU(6)_{\rm fs}$ representations of three
quarks are shown in Fig.~(\ref{fig4}) in terms of Young Diagrams.  The
$56_{\rm fs}$ is totally symmetric in exchange of spin and flavor, the
$70_{\rm fs}$ has mixed symmetry, and the $20_{\rm fs}$ is totally
antisymmetric.  The $56_{\rm fs}$ and $70_{\rm fs}$ are prominent in
the baryon spectrum.  For example, the lightest baryons form a
$56_{\rm fs}$ containing the decuplet with spin-$\fract{3}{2}$ and the
octet with spin-$\half$.  In fact \emph{all known baryons can be
assigned to either $56_{\rm fs}$ or $70_{\rm fs}$ representations}. 
What happened to the $20_{\rm fs}$?  This representation goes with a
totally antisymmetric space wavefunction, and could occur among the
positive parity baryon resonances, which are quite well studied.  The
$SU(3)_{\rm f}\times$spin content of the $20_{\rm fs}$ is $|8_{\rm
f}2_{\rm s}\rangle$ and
$|1_{\rm f}4_{\rm s}\rangle$.  This is the only occurrence of a
spin-$\fract{3}{2}$, $SU(3)_f$ singlet.  So observation of a spin
$\fract{3}{2}-\Lambda^{\ast 0}$ without the partners required to form
an octet would be the signature of the $20_{\rm fs}$.  There is no
serious candidate for a $1_{\rm f}4_{\rm s}$ state in the baryon
spectrum.  The QCD Breit interaction predicts this.  Two quarks in the
$20_{\rm fs}$ must be antisymmetric in simultaneous exchange of
flavor and spin.  In contrast, quark pairs in the $56_{\rm fs}$ must,
and in the $70_{\rm fs}$ may, be symmetric in flavor$\times$spin.  So
the $20_{\rm fs}$ alone decouples from the most attractive
configuration, $|\bar 3_{\rm f}1_{\rm s}\bar 3_{\rm c}\rangle$.  Because of this,
the $20_{\rm fs}$ is promoted to higher energy and is less stable.

%
%

\section{Mesons}
\label{section4}
The generalization of the Breit interaction to $\bar q q$ is simple. 
Once $\bar q q$ form a color singlet the residual interaction is
attractive for spin-$0$ and repulsive for spin-$1$, much like
positronium.  The consequences are well known: The pseudoscalar mesons
like the $\pi$ are lighter than their vector partners like the~$\rho$.

The fact that colorspin has other significant consequences for meson 
spectroscopy is less well known.

\subsection{The Absence of Exotics}

The absence of $SU(3)_{\rm f}$ representations other than $1_{\rm f}$ and
$8_{\rm f}$ for mesons, and $1_{\rm f}$, $8_{\rm f}$, and $10_{\rm f}$ for baryons was
one of the original inspirations for Gell-Mann's and Zweig's quark
model.  Later, QCD and confinement explained the absence of
representations of with fractional charge, but quark model exotics
like a meson $27_{\rm f}$-plet ($\bar q^{2}q^{2}$) or a baryon
$\overline{10}_{\rm f}$-plet ($q^{4}\bar q$) are allowed by confinement. 
Gell-Mann's original paper on the quark model begs the question with a
breathtakingly appropriate typographical error:

\begin{quote}
\noindent Baryons can now be constructed from quarks by using the 
combinations ($qqq$), ($qqqq\bar q$), etc., while mesons are made out 
of ($q\bar q$), ($qq\bar q\bar q$), etc.  It is assuming [{\it sic\/}] 
that the lowest baryon configuration ($qqq$) gives just the 
representations {\bf 1}, {\bf 8}, and {\bf 10} that have been 
observed, while the lowest meson configuration ($q\bar q$) similarly 
gives just {\bf 1} and~{\bf 8}.\cite{gellmann}
\end{quote}
and almost everyone has been assuming it ever since.

The experimental situation has changed little since Gell-Mann's day.  
Consider, for example, states of two pions at low energies.  The 
$I=2$ state, exemplified by $\pi^{+}\pi^{+}$ is exotic.  It contains 
at least $\bar d^{2}u^{2}$.  Data on $\pi^{+}\pi^{+}$ scattering 
shows no resonance at low energies, only a weak repulsion.  
The $I=1$ and $I=0$ states couple to $\bar q q$.  The $I=1$ phase 
shift resonates at the $\rho$.  The $I=0$ phase shift shows a strong, 
broad attraction from threshold up through $m_{\pi\pi}\approx 1$ GeV\null.
The same absence of striking resonances occurs in every exotic channel.

QCD is consistent with the absence of strongly bound mesons made of 
more than $\bar q q$.  Confinement saturates at $q^3$ or $\bar q q$.  
There are no van der Waals forces in QCD because there are no massless 
hadrons.  So the mechanism that forms molecules from atoms in QED is 
absent in QCD\null.  However, this does not explain why no $\bar 
q^{2}q^{2}$ states have been seen.

The QCD Breit interaction, ${\cal H}_{\rm eff}$, gives a simple
explanation for the absence of exotics.  A complete treatment would
require me to include $\bar q q$ interactions as well as $qq$. 
However, the gist of the argument can be read off Table~\ref{tab1}. 
Exotic $SU(3)_{\rm f}$ representations require either the quarks to be in
the $6_{\rm f}$ or the antiquarks to be in the $\bar 6_{\rm f}$ representation
or both.  If the quarks are in the $\bar 3_{\rm f}$ and the antiquarks are
in the $3_{\rm f}$, then the resulting $\bar q^{2} q^{2}$ state
transforms as an $8_{\rm f} \oplus 1_{\rm f}$, which is not exotic and could
 mistakenly be interpreted as a $\bar q q$ multiplet.  Table
\ref{tab1} shows that colorspin interactions are repulsive in the
$6_{\rm f}$ and $\bar 6_{\rm f}$ representations.  So colorspin pushes exotic
flavor states of $\bar q^{2} q^{2}$ up above thresholds where they
fall apart into the light pseudoscalar mesons from which they are
built.  A more sophisticated treatment including $\bar q q$ colorspin
forces confirms this result.\cite{Jaffe:1977ig}

So the QCD Breit interaction explains the absence of prominent exotic
meson resonances and also suggests that we look carefully at ordinary
mesons to see if any $\bar q^{2} q^{2}$ states have been mistakenly
cataloged as $\bar q q$.

\subsection{The Scalar Mesons}

The lightest (negative parity) light quark mesons have been known for
forty years.  The first excited states (with positive parity) are also
well known.  Of these, only the scalar mesons remain controversial.  The rest
were discovered and classified during the sixties and seventies.  The
identification and nature of the scalar mesons continues to be one of
the most controversial subjects in hadron spectroscopy.  The special
role of the $\pi\pi$ s-wave in chiral dynamics and in nuclear physics
makes this an important issue.

To make $J^{PC}=0^{++}$ in the naive quark model it is necessary to
add a unit of orbital angular momentum to a $\bar q q$ pair (to get
positive parity).  Apparently this costs more than half a GeV in mass,
since well established mesons with this makeup ($1^{++}, 1^{+-}$, and
$2^{++}$) are all known to lie between $1.2$ and $1.6$~MeV\null. In
contrast $\bar q^{2} q^{2}$ can couple to $0^{++}$ without excitation. 
Surprisingly there are striking experimental effects in all flavor
nonet $0^{++}$ channels at masses below 1 GeV, and they do not fit
naive $\bar q q$ quark model expectations.  Recently Black et al.\
reviewed the situation and concluded that the light $0^{++}$ mesons
look more like $\bar q^2 q^2$ than $\bar q q$, an idea originally
suggested on the basis of the QCD colorspin interactions back in the
1970's.\cite{Jaffe:1977ig,Black:1998wt}

The QCD Breit interaction between quarks strongly favors the $|\bar
3_{\rm f}1_{\rm s}\bar 3_{\rm c}\rangle$ configuration for quark pairs.  The $\bar
q q $ interaction, which we have not discussed, mixes in a significant
amount of the second most attractive configuration, $|\bar 3_{\rm f}3_{\rm s}
6_{\rm c}\rangle$, which, notice, is also a flavor $\bar 3_{\rm f}$.  Thus the
lightest $\bar q^2 q^2 $ mesons have the flavor structure $(\bar
q^{2})^{3_{\rm f}}(q^{2})^{\bar 3_{\rm f}}$, which leads to an octet plus a
singlet ($\equiv$ a nonet).  Diagonalization of the $\bar q^2 q^2
$ generalization of ${\cal H}_{\rm eff}$ in the $\bar q^2 q^2$
sector drives a single $0^{++}$ flavor nonet to remarkably low mass
($600$--$1000$~MeV in the Bag Model).  Thus the \emph{only} light meson
channel that has defied classification for the past forty years is
the one in which colorspin considerations leads one to expect $\bar q
\bar q q q$ to be important.

\begin{figure}[t]
\begin{center}
\BoxedEPSF{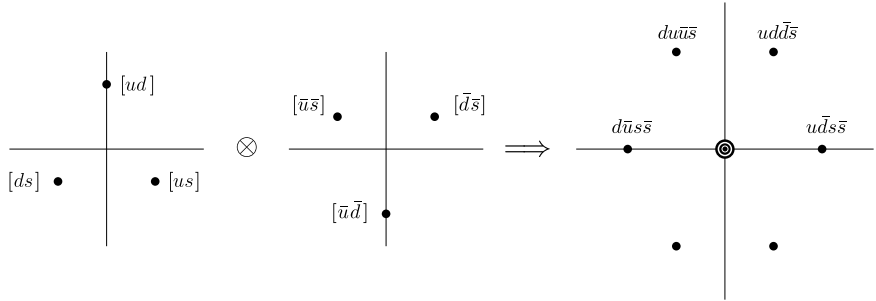 scaled 750} 
\end{center}
\caption{$SU(3)_{\rm f}$ weight diagrams for two antisymmetrized quarks,
two antisymmetrized antiquarks, and for the flavor nonet contructed from
them.\vspace*{-1pc}}
\label{fig5}
 \end{figure}

\begin{figure}[t]
\begin{center}
\BoxedEPSF{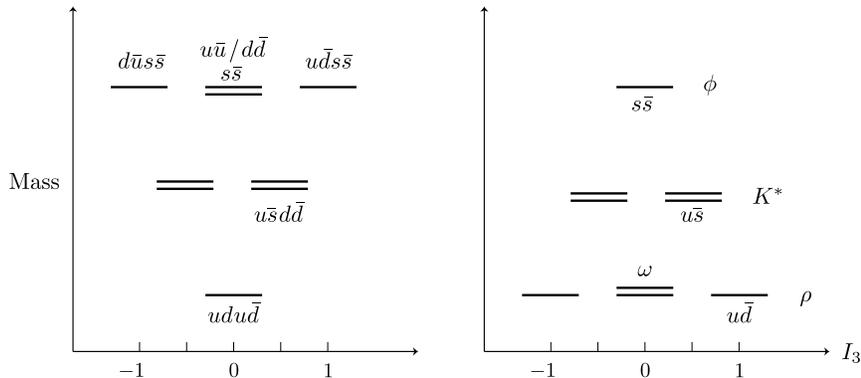 scaled 875} 
\end{center}
 \caption{A cartoon representation of the masses of a $\bar q \bar q q q$ nonet 
compared with a $\bar q q $ nonet.\vspace*{-1pc}}
\label{fig6}
 \end{figure}

Figure (\ref{fig5}) shows how a nonet can be constructed from two
quarks in a ${\bar 3}_{\rm f}$ and two antiquarks in a ${ 3}_{\rm f}$.  The
quark flavor content of some of the states are shown in the figure. 
The most striking feature of a $\bar q^2 q^2 $ nonet is an
\emph{inverted mass spectrum}.  A standard, magically-mixed $\bar q q$
nonet has a degenerate isosinglet and isovector at the bottom, a
strange isodoublet in the middle, and a lone isosinglet at the top. 
This ordering is easily understood by counting the number of strange
quarks in the meson.  A $\bar q^2 q^2$ nonet is opposite: the
$\bar q^2 q^2 $ isotriplet and one of the isosinglets contain hidden
strange quarks,
 $\{u\bar d s \bar s, \frac{1}{\sqrt{2}}(u\bar u-d\bar
d)s\bar s, d\bar u s \bar s\}$ and $\frac{1}{\sqrt{2}}(u\bar u+d\bar
d)s\bar s$, and therefore lie at the top of the multiplet.  The
other isosinglet, $u\bar d d \bar u$ is the only state without strange
quarks and therefore lies alone at the bottom of the multiplet.  The
strange isodoublets should lie in between.  In summary, one expects a
degenerate isosinglet and isotriplet at the top of the multiplet and
strongly coupled $\bar K K$, an isosinglet at the bottom, strongly
coupled to $\pi\pi$, and a strange isodoublet coupling to $K\pi$ in
between.  Fig.~(\ref{fig6}) shows the mass spectrum and quark content
of a $\bar q^2 q^2 $ nonet and, for comparison, a $\bar q q$
nonet like the vector mesons ($\rho$, $\omega$, and $\phi\quad K^*$).

The most well established $0^{++}$ mesons are the $I=0$ $f_{0}(980)$ 
and the $I=1$ $a_{0}(980)$.  The $f_{0}$ couples to
$\pi\pi$ and $\bar K K$.  The $a_{0}$ couples to $\pi\eta$ and 
$\bar K
K$.  Both are so close to the $\bar K K$ threshold at 990~MeV that
their shapes are strongly distorted by threshold effects, but it is
clear that they couple strongly to $\bar K K$.  This has always
troubled those who would like to identify them as $\bar q q$ states. 
It is a natural consequence of the hidden $\bar s s$ component if they
are predominantly $\bar q\bar q q q$.\footnote{Another interpretation 
of the $f_{0}$ and $a_{0}$ as ``$\bar K K$ molecules'' is closely 
related to the $\bar q\bar q q q $ interpretation.\cite{Isgur}}

The contribution of Black et al.\ centers on the other isosinglet and
strange isodoublet needed to fill up a nonet.  If they are
predominantly $\bar q^2 q^2$ states, they should couple strongly
to $\pi\pi$ and $K \pi$ respectively.  Since the $\pi\pi$ and $K\pi$
thresholds are very low, one expects these states to ``fall apart'' into
$\pi\pi$ or $K\pi$ with very large width.  Enhancements in $\pi\pi$
and $K\pi$ s-wave scattering have been known for decades.  Black et al.\ 
make the case that these enhancements correspond to broad states at
approximately 560~MeV in $\pi\pi$ (known as the $\sigma$(560)) and at
900~MeV in $K\pi$ (known as the $\kappa(900)$).  Together the
$f_{0}$(970), $a_{0}$(980), $\sigma$(560), and $\kappa$(900) make a
nonet with mass spectrum, decay couplings and widths that look
qualitatively like a $\bar q\bar q q q $ system.  Should this
assignment hold up, it will be striking confirmation of the role of
the QCD Breit interaction in hadron spectroscopy.

\section{Back to Basics:  Spectroscopic Rules for $\mathbf{Q^{N}}$}
\label{section5}
In Section~\ref{section2} I looked at the colorspin force between pairs of
quarks.  Here I will generalize that analysis to the case of $N$
quarks all in the same spacial orbital.  This allows us to look at
systems of up to 18 quarks (3 colors $\times$ 3 flavors $\times$ 2
spins) at short distances.  It will allow us to draw interesting
qualitative conclusions about baryons and ``dibaryons'' (i.e.,  6-quark
systems).

For $N$ quarks,
\begin{equation}
        {\cal H}_{\rm eff}^{N} \propto -\sum_{i\ne j}^{N}
        \{\undertilde{\lambda}\vec{\sigma}\}_{i}
		\cdot\{\undertilde{\lambda}\vec{\sigma}\}_{j}\, .
        \la{4.0}
\end{equation}
Here I have grouped the color and spin operators of the i$^{\rm th}$
quark together.  The 24 matrices $\{\undertilde{\lambda}\vec\sigma\}$
together with the 3 Pauli matrices $\{\vec\sigma\}$ and the 8
Gell-Mann matrices $\{\undertilde{\lambda}\}$ together form the 35
generators of an $SU(6)_{\rm cs}$ symmetry that we can call
``colorspin''.  Let us denote the colorspin generators as a
35-dimensional vector of 6$\times$6 matrices (in the fundamental
representation), $\{\mu^{r},r=1,\dots,35\}$.  The quadratic Casimir operator
of $SU(6)_{\rm cs}$ is defined as the sum of the squares of the colorspin
generators,
\begin{equation}
	C_{6}^{N} = \sum_{r=1}^{35}\Bigl(\sum_{i=1}^{N} 
	\mu^{r}_{i}\Bigr)^{2}
	\la{4.1}
\end{equation}
in analogy to the total spin
\begin{equation}
	C_{2}^{N} = 4S_{N}(S_{N}+1)=\sum_{k=1}^{3}\Bigl(\sum_{i=1}^{N} 
	\sigma^{k}_{i}\Bigr)^{2}
	\la{4.2}
\end{equation}
and total color,
\begin{equation}
	C_{3}^{N} =\sum_{a=1}^{8}\Bigl(\sum_{i=1}^{N} 
	\lambda^{a}_{i}\Bigr)^{2}.
	\la{4.3}
\end{equation}
Of course the color Casimir, $C_{3}^{N}$, is zero for any physical, i.e., 
color singlet, hadron.

It is an elementary exercise in matrix algebra to rewrite 
${\cal H}_{\rm eff}^{N}$ in terms of $N$, $S_{N}$, and $C_{6}^{N}$,
\begin{equation}
	{\cal H}_{\rm eff}^{N} \propto 8N-\hf C_{6}^{N} +\fract{4}{3}
	S_{N}(S_{N}+1)\, .
	\la{4.4}
\end{equation}

The Casimir of SU(6)$_{\rm cs}$ is the sum of 35 normed generators.  
$S_{N}$ is the sum of only 3 identically normed generators.  
Therefore $C_{6}^{N}$ dominates over $S_{N}$ in \eq{4.4}.
So the spectroscopic consequences of ${\cal H}_{\rm eff}^{N}$ for $Q^{N}$ 
states are very simple.  It is important to remember that we are 
assuming that all quarks are in the same orbital of some mean 
field and that their masses can be ignored.\footnote{Ignoring the 
$s$-quark mass is not a good approximation.  Fortunately $s$-quark 
mass effects can be calculated perturbatively.}  
\bit
	\item 
	In this approximation, the Casimirs of spin, flavor, color and colorspin
	all commute with ${\cal H}_{\rm eff}^{N}$, so mass eigenstates are
	(color singlet) states in irreducible representations of 
	$SU(3)_{\rm f}$, $SU(6)_{\rm cs}$, and spin.
	\item
	For a given number of quarks, the largest $C_{6}^{N}$ is associated
	with the most symmetric representation -- the most ``horizontal''
	Young Diagram.  Since quarks in the same orbital must be
	antisymmetrized in color$\times$spin$\times$flavor, the most
	symmetric colorspin goes with the most antisymmetric flavor.  Thus the
	lightest state of $N$ quarks is the one \emph{most antisymmetric in
	flavor} subject to the constraint that the associated (conjugate)
	colorspin representation must contain a color singlet.
	\item 
	The factor of $S_{N}(S_{N}+1)$ is generally too small to affect 
	the order of states.
\eit
In short:  Nature prefers flavor antisymmetry.  It is illuminating to 
apply this principle to the well-known case of $N=3$ and then to the 
more speculative case of $N=6$.

\section{Baryons and Dibaryons}
\label{section6}
The simple rule for the spectrum of $N$ quarks has interesting 
consequences for $N=6$.  First, however, let us see how it works for 
the baryons, $N=3$.

\subsection{Baryons Reconsidered}

\begin{figure}[t]
\begin{center}
\BoxedEPSF{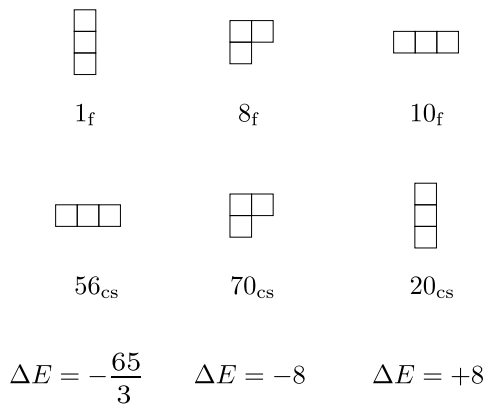} 
\end{center}
 \caption{The states of three quarks, classified in $SU(6)_{\rm cs}\otimes 
SU(3)_{\rm f}$.  For each multiplet the colorspin and flavor Young Diagrams are 
shown.  To make states antisymmetric under flavor$\times$color$\times$spin, 
quarks must have the conjugate permutation symmetry under flavor and 
colorspin, obtained by interchanging rows and columns in the Young 
Diagram.  The Breit Interaction favors the most antisymmetric flavor 
configuration, as evidenced by the eigenvalues displayed with the
diagrams.\vspace*{-1pc}}
\label{fig7}
 \end{figure}

The allowed colorspin and flavor-irreducible representations are
tabulated in terms of Young Diagrams in Fig.~(\ref{fig7}).  The flavor
$SU(3)_{\rm f}$ representations,~${1}_{\rm f}$, ${  8}_{\rm f}$, and~${10}_{\rm
f}$ are paired with the colorspin $SU(6)_{\rm cs}$  representations $56_{\rm
cs}$, $70_{\rm cs}$, and $20_{\rm cs}$,  respectively.  The ``flavor
antisymmetry'' rule dictates 
\begin{equation}
	M_{  1} < M_{  8} <M_{  10}\ .
	\la{5.1}
\end{equation}
In fact the expectation values of ${\cal H}^{N=3}_{\rm eff}$ in the
three states are $\langle{\cal H}_{\rm eff}\rangle_{  1} =
-65/3$, $\langle{\cal H}_{\rm eff}\rangle_{  8} = -8$, and
$\langle{\cal H}_{\rm eff}\rangle_{  10} = +8$.  To complete the
classification we must find out which of these colorspin multiplets
contain color singlets.  The $SU(3)_{\rm c}\times SU(2)_{\rm s}$ content of 
$SU(6)_{\rm cs}$
representations is best known in the context of Gursey's old
$SU(6)_{\rm fs}$.  We can copy those results over for color$\times$spin:
\begin{eqnarray}
      	56_{\rm cs} &\supset& {  8}_{\rm c}2_{\rm s}\oplus
	{  10}_{\rm c}4_{\rm s}\nonumber\\
	70_{\rm cs} &\supset& {  1}_{\rm c}2_{\rm s}\oplus
	{  8}_{\rm c}2_{\rm s}\oplus
	{  8}_{\rm c}4_{\rm s}\oplus
	{  10}_{\rm c}2_{\rm s}\nonumber\\
	20_{\rm cs} &\supset& {  1}_{\rm c}4_{\rm s}\oplus
	{  8}_{\rm c}2_{\rm s}
	\la{5.2}
\end{eqnarray}

 A novel perspective on the QCD picture of baryons emerges.  The most
attractive colorspin channel, which would be a
 candidate for the lightest baryon, the flavor singlet $56_{\rm cs}$,
 is not a physical state because it does not contain a
 color singlet.  So the colorspin rule to antisymmetrize flavor is, in
 this case, frustrated by Fermi statistics: a color singlet, flavor
 singlet three quark state would have to be antisymmetric in spin,
 which is not possible for spin-$\half$.  Otherwise, the baryon
 ground state would be a flavor singlet, $\Lambda$-like $uds$ baryon.

The flavor-antisymmetry rule works fine for the rest of the three
quark states; the mixed symmetry multiplet, the $70_{\rm cs}$, is
lightest.  It contains a color singlet with spin-$\half$ -- the
familiar flavor octet baryons.  The antisymmetric colorspin multiplet, the
$20_{\rm cs}$, is heaviest.  It contains a color singlet with
spin-$\fract{3}{2}$, the familiar decuplet baryons.

One final comment: It is tempting to suggest that the excluded 
multiplet, the $56_{\rm cs}$, could neutralize its color with a 
constituent gluon.  The result would be a $qqqg$ $SU(3)_{\rm f}$ singlet 
``hybrid'' baryon.  There is a rather light neg\-ative-parity $\Lambda$ 
at 1405~MeV that has often caused problems for quark modelers.  
However current wisdom favors a $q^{3}$ or $q^{4}\bar q$ 
interpretation.\cite{Pakvasa:1999zv}

\subsection{Dibaryons -- Known and Unknown}

\begin{figure}[t]
\begin{center}
\BoxedEPSF{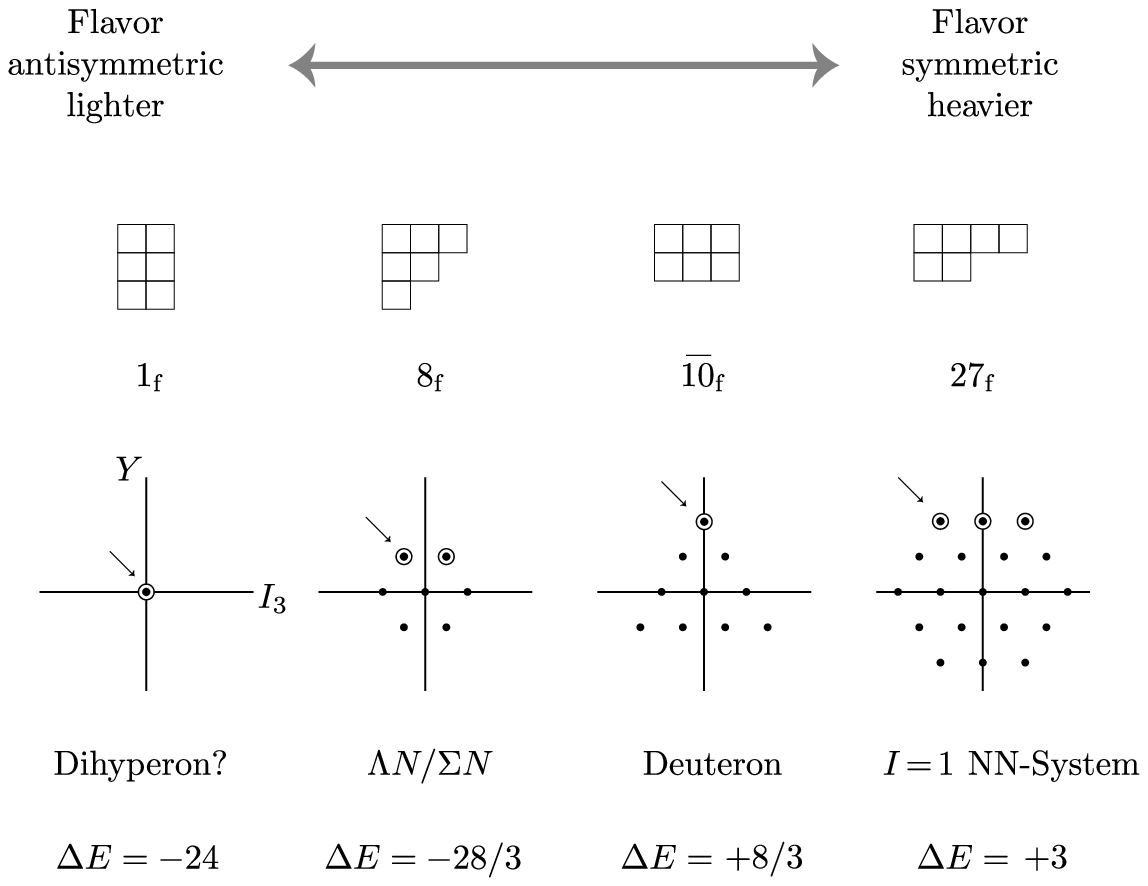} 
\end{center}
 \caption{Some of the states of six quarks, classified in $SU(6)_{\rm cs}\otimes 
SU(3)_{\rm f}$.  Only the flavor Young Diagrams are shown.  The colorspin 
diagrams are conjugate.  Again, the most antisymmetric possible flavor 
state has the most attractive colorspin interaction.  The multiplets shown 
are interesting because the correspond to physically interesting channels: 
the $H$, the $\Lambda N/\Sigma N$ system, the deuteron, and the isotriplet 
dinucleon, respectively.\vspace*{-1pc}}
\label{fig8}
\end{figure}

Now consider $N=6$. The flavor Young Diagrams for six quarks are 
shown in Fig.~(\ref{fig8}).  The colorspin Young Diagrams are conjugate (rows 
and columns interchanged).  Only a few of the multiplets are shown.  
For the multiplets shown, the colorspin interaction energies are $\langle{\cal
H}_{\rm eff}\rangle_{1} = -24$, 
$\langle{\cal H}_{\rm eff}\rangle_{  8} = -28/3$, $\langle{\cal 
H}_{\rm eff}\rangle_{\overline{10}} = +8/3$, and $\langle{\cal 
H}_{\rm eff}\rangle_{27} = +3$. 

The most attractive channel is the flavor singlet ``dihyperon'' (known 
as the $H$) with quark content $u^{2}d^{2}s^{2}$.\cite{Jaffe:1977yi} 
If colorspin were the only consideration, its interaction energy, 
$-24$, would bind it with respect to decay into two octet baryons 
(colorspin energy $2\times-8$).  It would decay into the would-be 
flavor singlet baryon (colorspin energy $2\times -65/3$), if that state 
were not excluded by color confinement.  Models of confined light 
quarks are not sufficiently accurate to decide unequivocally whether 
the $H$ is bound.  Lattice calculations are improving and may answer 
the question before experimenters are able either to discover it or 
establish that it is not bound.\cite{Pochinsky:1998zi}

The next most attractive channel is the flavor octet.  The hypercharge
$+1$ members have the quantum numbers of $\Sigma N$ or $\Lambda N$.  A
strong resonance with these quantum numbers has long been known in
$\Lambda N$ scattering at 2129~MeV\null. Unfortunately this cannot be
unambiguously interpreted as evidence for colorspin attraction because
a loosely bound state with the same quantum numbers is expected as an
$SU(3)_{\rm f}$ analog of the deuteron.

The ${ \overline{10}}_{\rm f}$ and ${ 27}_{\rm f}$ channels are interesting
because the colorspin interaction energy is repulsive.  These flavor
representations are the smallest ones that contain states with the
quantum numbers of the deuteron (the ${\overline{ 10}}_{\rm f}$) and the
$I=1$ two nucleon system (the ${ 27}_{\rm f}$).  Both appear at $Y=2$
(which is strangeness zero) in the $SU(3)_{\rm f}$ weight diagrams shown
in Fig.~(\ref{fig8}).  I interpret the positive colorspin interaction energy as
evidence that the quark Breit interaction is repulsive at short
distances in these channels.  Once the three quark baryons have
separated, apparently other effects generate an intermediate range
attraction, which binds the deuteron and almost binds the $I=0$ $NN$
system.

\section{Colorspin in Quark Matter}
\label{section7}
The most interesting recent application of colorspin systematics in
QCD is the study of condensates and symmetries in cold, dense
matter.\cite{Alford:1997zt} The attraction between quark-antiquark
pairs is so strong in QCD that $\bar q q$ pairs condense in the vacuum,
breaking chiral symmetry.  Calculations show that the color singlet,
spin-zero (${ 1}_{\rm c}1_{\rm s}$) configuration is the most attractive
channel for ${\cal H}_{\rm eff}$.  So condensation in this channel is
consistent with colorspin considerations.  The $\bar q q$ condensate
breaks the flavor $SU(3)_{fL}\times SU(3)_{fR}$ chiral symmetry
spontaneously down to vector $SU(3)_{\rm f}$, the standard flavor symmetry of the
hadron spectrum.  Interestingly, the colorspin interaction energy in the
color antitriplet, spin-zero ($\bar 3_{\rm c}1_{\rm s}$) $qq$ channel is half as
attractive.  But no condensate develops in this channel in vacuo.  It
is forbidden by the Vafa-Witten Theorem, which says that vector symmetries
cannot break spontaneously in vector gauge theories like QCD\null.\cite{VW}

Things change in the presence of matter.\cite{BL,Alford:1997zt} At
zero temperature and finite density, degeneracy effects become
important.  At some point close to where the chemical potential for
baryon number ($\mu_{B}$) becomes comparable to $\Lambda_{QCD}$,
nuclear matter passes over to quark matter.  Studies show that the
attraction that drives $\bar q q$ condensation weakens as $\mu_{B}$
increases.  They also show that the $qq$ colorspin attraction in the
$\bar 3_{\rm c}1_{\rm s}$ channel can generate a condensate in the vicinity of
the Fermi surface similar to BCS pairing in superconductivity. 
Alford, Rajagopal and Wilczek argue that this ``color
superconductivity'' is robust -- it occurs both in colorspin models of
the quark force and in instanton models as well.  Of course, QCD at
moderate densities is still strongly interacting, and these
predictions must be regarded as qualitative.  Authoritative statements
await a lattice treatment of quark matter at finite chemical
potential.

If $q q$ condenses in the ${\bar 3}_{\rm c}1_{\rm s}$ channel in 
quark matter, color symmetry is broken from $SU(3)_{\rm c}$ to 
$SU(2)_{\rm c}$.  Of the eight gluons, the three generators of 
color-$SU(2)_{\rm c}$ remain massless.  The four gluons that are doublets 
under $SU(2)_{\rm c}$ get mass a la the Meissner effect.  
One linear combination of the eighth (color hypercharge) gluon and the 
photon stays massless (much as the standard model photon survives 
weak $SU(2)$ symmetry breakdown).   $SU(2)_{\rm c}$ nonsinglet quark or 
gluon excitations still experience long-range confining forces, but 
$SU(2)_{\rm c}$ singlet states do not.  Thus, for example, two colors of each 
quark flavor (the ones that form the $SU(2)_{\rm c}$ doublet) remain 
confined, but the third color, say ``blue'', is a freely propagating 
fundamental excitation in the medium.

 For flavor 
symmetries, $qq$ condensation has interesting consequences.  The two-flavor
and three-flavor cases differ  dramatically.  In two-flavor QCD (obtained by
imagining the strange  quark mass to be much larger than $\mu_{B}$) the ${ 
\bar  3}_{\rm c}1_{\rm s}$ condensate is a  $SU(2)_{fL}\times SU(2)_{fR}$
singlet  so it has no effect on the flavor symmetries (though it does violate
baryon number).  In particular, chiral symmetry is unbroken.  Since 
chiral symmetry is broken spontaneously in the QCD vacuum, there must 
be a phase transition between $\mu_{B}=0$ and the high density phase 
where superconductivity sets in.

In three-flavor QCD (imagine the strange quark mass to be negligible 
compared to $\mu_{B}$) the $qq$ condensate is in the 
antisymmetric $\bar 3_{\rm c}1_{\rm s}\bar 3_{\rm f}$ channel.  Thus the 
condensate breaks both color and flavor $SU(3)$ symmetries.  However, 
the condensate is invariant under \emph{simultaneous} rotations in 
color and flavor, which means that a ``diagonal'' vector $SU(3)$ symmetry 
remains unbroken.  This is the same flavor structure as the low 
density phase (only a vector $SU(3)_{\rm f}$ remains unbroken in vacuo with three 
massless quarks), so it is possible that there is no phase 
transition from the low density to high density phase.  Dubbed 
``color-flavor locking'',\cite{CFL} this phenomenon leads to several 
interesting speculations on the phase structure of QCD at finite 
densities.  Implications for observations at RHIC and for the 
behavior of neutron stars are the subject of much recent 
interest.\cite{SRS,neutron}

With this I've come to the end of this tour of the phenomenological 
implications of the Breit interaction in QCD\null.  Insights into 
QCD in the confining domain are precious.  The underlying Lagrangian 
is so simple, the phenomena are so rich, and the theoretical 
structure is so complex.  It is remarkable that so much can be 
understood qualitatively in terms of the simple effective 
interaction,
\begin{equation}
        {\cal H}_{\rm eff} \propto\, -\!\sum_{i\ne j}
        {\undertilde{\lambda}}\,_{i}\cdot {\undertilde{\lambda}}\,_{j}
        \vec{\sigma}_{i}\cdot\vec{\sigma}_{j}
        \la{last}
\end{equation}
whose origins go back to Gregory Breit in the first half of the 
twentieth century.

\section*{Acknowledgments}
This work is supported in part by funds provided by the U.S.
Department of Energy (D.O.E.) under cooperative
research agreement \#DF-FC02-94ER40818.


\begin{thebibliography}{99}

\bibitem{DeRujula:1975ge}
A.~De Rujula, H.~Georgi and S.L.~Glashow,
\Journal{\PRD}{12}{147}{1975}. 

\bibitem{DeGrand:1975cf}
T.~DeGrand, R.L.~Jaffe, K.~Johnson and J.~Kiskis,
\Journal{\PRD}{12}{2060}{1975}. 

\bibitem{Jaffe:1977ig}
R.~Jaffe,
\Journal{\PRD}{15}{267}{1977};  
\Journal{\PRD}{15}{281}{1977}. 

\bibitem{Jaffe:1977yi}
R.~Jaffe,
\Journal{\PRL}{38}{195}{1977}.  

\bibitem{Black:1998wt}
D.~Black, A.H.~Fariborz, F.~Sannino, and J.~Schechter,
\Journal{\PRD}{59}{074026}{1999};  
hep-ph/9808415.

\bibitem{Alford:1997zt}
M.~Alford, K.~Rajagopal, and F.~Wilczek,
\Journal{\PLB}{422}{247}{1998};  
hep-ph/9711395.

\bibitem{Caso:1998}
C.~Caso et al.,
\Journal{\it Eur.\ Phys.\ J.\ \rm C}{3}{1}{1998}. 

\bibitem{Dalitz}
R.H.~Dalitz, ``Symmetries and the Strong Interactions'', in 
\emph{Proceedings of the XIIIth International Conference on High 
Energy Physics} (University of California Press, Berkeley, 1967); 
reprinted in J.J.J.~Kokkedee, \emph{The Quark Model} 
(W.A.~Benjamin, New York, 1969).

\bibitem{Gursey}
F.~G\"ursey and L.A.~Radicati,
\Journal{\PRL}{13}{173}{1964}. 

\bibitem{gellmann}
M.~Gell-Mann, \Journal{\it Phys.~Lett.}{8}{214}{1964}.

\bibitem{Isgur}N. Isgur and J. Weinstein, 
\Journal{\PRD}{41}{2236}{1990}.

\bibitem{Pakvasa:1999zv}
S.~Pakvasa and S.F.~Tuan,
\Journal{\PLB}{459}{301}{1999};  
hep-ph/\goodbreak9903551.

\bibitem{Pochinsky:1998zi}
A.~Pochinsky, J.W.~Negele, and B.~Scarlet,
\Journal{\it Nucl.\ Phys.\ Proc.\ Suppl.}{73}{255}{1999};  
 hep-lat/9809077.

\bibitem{VW}
C.~Vafa and E.~Witten, \Journal{\NPB}{234}{173}{1984}.

\bibitem{BL}
D.~Bailin and A.~Love, 
\Journal{\it Phys.\ Rept.}{107}{325}{1984}, 
and references therein. 
 
\bibitem{CFL}
M.~Alford, K.~Rajagopal, and F.~Wilczek,
\Journal{\NPB}{537}{443}{1999};  
 hep-ph/9804403.

\bibitem{SRS} M.~Stephanov, K.~Rajagopal, and 
E.~Shuryak,
\Journal{\PRL}{81}{4816}{1998}; 
hep-ph/9806219;
\Journal{\PRD}{60}{114028}{1999}; 
hep-ph/9903292.

\bibitem{neutron}
M.~Alford, J.~Berges, and K.~Rajagopal, hep-ph/9910254.

%
%
%

\end{thebibliography}
\end{document}